\documentclass{article}

\usepackage{arxiv}

\usepackage[utf8]{inputenc} 
\usepackage[T1]{fontenc}    
\usepackage{hyperref}       
\usepackage{url}            
\usepackage{booktabs}       
\usepackage{amsfonts}       
\usepackage{nicefrac}       
\usepackage{microtype}      
\usepackage{lipsum}		
\usepackage{graphicx}
\usepackage{doi}
\usepackage{amsmath}
\usepackage{authblk}
\usepackage[title]{appendix}
\usepackage[sort&compress,numbers]{natbib}

\title{Unsteady solutions of the spray flamelet equations}

\author[1]{Felipe Huenchuguala}
\author[1]{Francisco Rivadeneira}
\author[2]{Arne Scholtissek}
\author[2]{Christian Hasse}
\author[3]{Eva Gutheil}
\author[ ~,1]{Hernan Olguin\thanks{Corresponding author\\ Email address: hernan.olguin@usm.cl}}

\affil[1]{Department of Mechanical Engineering, Universidad Técnica Federico Santa María,\linebreak Avenida España 1680, Valparaíso, Chile\linebreak}
\affil[2]{Institute for Simulation of reactive Thermo-Fluid Systems, TU Darmstadt,\linebreak Otto-Berndt-Stra{\ss}e 2, 64287 Darmstadt, Germany\linebreak}
\affil[3]{Interdisciplinary Center for Scientific Computing, Heidelberg University,\linebreak Im Neuenheimer Feld 205, 69120 Heidelberg, Germany}

\date{}

\begin{document}
\maketitle

\begin{abstract}
Solutions of the spray flamelet equations reported in the literature during the last decade have been limited to very specific situations presenting steady evaporation profiles only. In contrast, intrinsically unsteady interactions between the liquid and gas phases have received little attention so far. In this work, the spray flamelet equations are closed by means of a Lagrangian description of the liquid phase in mixture fraction space, which allows solving them for unsteady situations. The resulting formulation is then employed to conduct parametric analyses of the effects of initial droplet radius and velocity variations on ethanol/air non-premixed gas flamelets perturbed by sprays generated with different droplet injection strategies. Special emphasis is given to the differences between continuous and discontinuous droplet injection. The results illustrate how the latter can considerably increase the temperature and stability of flamelet structures, provided the spray parameters are appropriately selected.
\end{abstract}

\vspace{15mm}

\section{Introduction}\label{introduction}

In non-premixed gas flames, the time scales of chemical reactions are typically much shorter than the ones associated with convection and diffusion. As a consequence, combustion tends to take place in very thin layers admitting a one-dimensional description in terms of the mixture fraction, $Z$, which is formalized through the so-called flamelet equations. In their original version, derived by Peters~\cite{Peters84}, these equations present a single unclosed quantity: The scalar dissipation rate, $\chi = 2D_Z |\nabla Z|^2$, with $D_Z$ denoting a diffusion coefficient. This variable can be directly closed in mixture fraction space, either by means of an analytical expression, such as the inverse of the complementary error function~\cite{Peters84} or the ln-profile~\cite{Pitsch00}, or by the consideration of a transport equation~\cite{Hasse05,Scholtissek17,Olguin231}. In all cases, the only free parameter remaining is the strain rate, $a$, which can be directly varied from very low values (flames close to equilibrium) up to flame extinction, such that the entire solution space of the flamelet equations can be easily covered. This has allowed a systematic study of non-premixed flamelet structures and the development of a deep physical understanding of them, both in steady~\cite{Peters84,Pitsch98,Claramunt06,Carbonell09,Scholtissek15} and unsteady situations~\cite{Pitsch00,Hasse05,Claramunt06,Ghoniem92,Barlow92,Chan98,Mauss91,Barths00,Cuenot00,Pitsch03,Ihme05,Felsch09,Bekdemir11,Verhoeven12,Dhuchakallaya13,Hewson13,Ameen15,Sun2020}.

The big success of flamelet theory in the description of non-premixed combustion has motivated its extension to spray flames, which has led to the so-called spray flamelet equations~\cite{Luo13, Olguin14, Olguin142, Olguin144, Franzelli15, Olguin19}. While in appearance very similar to Peters equations~\cite{Peters84}, the inclusion of evaporation makes a systematic study of the entire solution space of these extended formulations very difficult. First, spray flamelet structures do not only depend on the strain rate, but also on the liquid-fuel to air equivalence ratio, $E$, the initial droplet velocity, $u_0$, and the initial droplet radius, $R_0$~\cite{Hollmann96,Gutheil98,Hollmann98,Ge081,Hu17,Hu172,Hu20}. Additionally, for a given set of parameters, ($a$, $E$, $R_0$, $u_0$), the droplet injection strategy will determine whether a steady (continuous injection) or a pseudo-steady (discontinuous injection) evaporation profile will be obtained. Because of these difficulties, analyses of the spray flamelet equations reported in the literature have been limited to very specific situations considering steady evaporation profiles only~\cite{Franzelli15, Olguin19}. As a consequence, big regions of their solution space remain unexplored and the possibility of  studying important, inherently unsteady two-phase interactions has been completely excluded. This is not a satisfactory state of affairs and represents a major motivation for the present research.

The \textbf{main objective} of this work is improving the current understanding of the spray flamelet equations by extending their analysis to regions of their solution space that have not been studied so far. More specifically, the formulation presented in~\cite{Olguin19}, extended to consider unsteady effects, is complemented by a Lagrangian description of the liquid phase in composition space. With the resulting set of equations, non-premixed ethanol/air gas flamelet structures perturbed by different sprays are analyzed, emphasizing the differences between continuous and discontinuous droplet injection strategies, which lead to steady and unsteady evaporation profiles, respectively. The comparison is focused on the main differences between the flamelets in terms of their i) structure, and ii) stability. This study is expected to provide an appropriate basis for further research on the unsteady behavior of the spray flamelet equations and the ways in which evaporation can be used to improve combustion processes.


\section{Mathematical Model}\label{sec:math}

In this section, the mathematical formulation to be employed for the description of the different flamelet structures of interest is introduced. This includes a summary of the spray flamelet equations (Section~\ref{sec:flam_eq}), and a closure model for the liquid phase in composition space (Section~\ref{sec:closure}). Additionally, the algorithm employed to solve the resulting system is explained in Section~\ref{sec:solstrat}.


\subsection{Spray Flamelet Equations}\label{sec:flam_eq}

The spray flamelet equations considered in this work are an unsteady extension of the formulation presented in~\cite{Olguin19}. For chemical species mass fractions and temperature these yield
\begin{align}
\rho \frac{\partial Y_k}{\partial\tau}=&\frac{\rho}{\mathrm{Le}_k}\frac{\chi}{2}\frac{\partial^2Y_k}{\partial Z^2}+\dot{\omega}_k 
-\dot{S}_v(1-Z)\frac{\partial Y_k}{\partial Z}+\dot{S}_v(\delta_{kF}-Y_k)+\Omega_{Y_k},
\label{eq:Yk}
\end{align}
and
\begin{align}
\rho C_p\frac{\partial T}{\partial\tau}=&\rho C_p\frac{\chi}{2}\frac{\partial^2T}{\partial Z^2}+\dot{\omega}_T 
-C_p\dot{S}_v(1-Z)\frac{\partial T}{\partial Z}+\dot{S}_e+\Omega_T,\label{eq:T}
\end{align}
respectively. Here, $\rho$ denotes the gas density, $C_p$ is the specific heat at constant pressure of the gas mixture and $D_Z$ is the diffusion coefficient of the mixture fraction, which is assumed to be equal to $\lambda/\rho C_p$ ($\mathrm{Le}_Z=1$), where $\lambda$ is the thermal conductivity and $\mathrm{Le}_k$ is the Lewis number of the $k$-species. Further, $\delta_{kF}$ represents the Kronecker delta, where the subscript $F$ refers to fuel. The mass and energy sources due to evaporation are denoted as $\dot{S}_v$ and $\dot{S}_e$, while $\dot{\omega}_k$ and $\dot{\omega}_T$ are the chemical reaction rate and the energy source term due to chemical reactions, respectively. Finally, $\Omega_{Y_k}$ and $\Omega_T$ comprise all effects that, while important for the exact definition of the spray flamelet structures, are expected to be small. They can be expressed as
\begin{align}
\Omega_{Y_k}=&-\sqrt{\frac{\chi}{2D_Z}}\frac{\partial}{\partial Z}\left(\rho D_Z\left(\frac{\mathrm{Le}_k-1}{\mathrm{Le}_k}\right)\sqrt{\frac{\chi}{2D_Z}}\right)\frac{\partial Y_k}{\partial Z} 
 -\sqrt{\frac{\chi}{2D_Z}}\frac{\partial(\rho \widetilde{V}_{kZ}Y_k)}{\partial Z},
\label{eq:omegaYk}
\end{align}
and
\begin{align}
\Omega_T=\left(\rho \frac{\chi}{2}\frac{\partial C_p}{\partial Z}-\rho \sqrt{\frac{\chi}{2D_Z}}\sum_{k=1}^N C_{pk}V_{kZ}Y_k\right)\frac{\partial T}{\partial Z},
\end{align}
where $C_{pk}$ is the specific heat at constant pressure of species $k$, and $V_{kZ}$ and $\widetilde{V}_{kZ}$ are diffusion velocities in mixture fraction space
\begin{align}
V_{kZ} = -\sqrt{\frac{\chi}{2D_Z}}\frac{D_k}{Y_k}\frac{\partial Y_k}{\partial Z}+\widetilde{V}_{kZ}
\end{align}
and
\begin{align}
\widetilde{V}_{kZ} = -\sqrt{\frac{\chi}{2D_Z}}\frac{D_k}{\overline{W}}\frac{\partial \overline{W}}{\partial Z}-\sqrt{\frac{\chi}{2D_Z}}\frac{D_{Tk}}{\rho TY_k}\frac{\partial T}{\partial Z}+{V}_{kZ}^C.
\end{align}
In these equations, $D_k$ and $D_{Tk}$ denotes the diffusion and thermal diffusion coefficients of species $k$ into the mixture, respectively, while $\overline{W}$ represent the mean molecular weight of the mixture. The velocity correction ensuring mass conservation, ${V}_{kZ}^C$, is
\begin{align}
V_{kZ}^C=\sqrt{\frac{\chi}{2D_Z}}\sum_{k=1}^N\left(\frac{D_kY_k}{X_k}\frac{\partial X_k}{\partial Z}+\frac{D_{Tk}}{\rho T}\frac{\partial T}{\partial Z}\right),
\end{align}
with $X_k$ denoting the mole fraction of species $k$.
 
For the closure of the scalar dissipation rate, $\chi$, a transport equation for the gradient of the mixture fraction, $g_Z = \lvert \nabla Z \rvert$, is considered, which has been shown to lead to a solvable and easy to handle system~\cite{Olguin19}. In particular, the following equation will be adopted
\begin{align}
\frac{\partial g_Z}{\partial\tau}=&g_Za+g_Z^2\frac{\partial}{\partial Z}\left(\frac{1}{\rho}\frac{\partial(\rho D_Zg_Z)}{\partial Z}\right)
+g_Z^2\frac{\partial}{\partial Z}\left(\frac{\dot{S}_v(1-Z)}{\rho g_Z}\right), \label{eq:gz}
\end{align}
which is the corresponding extension of the formulation presented in~\cite{Olguin19} accounting for unsteady variations. In Eq.~\eqref{eq:gz}, and given the assumption of a physical space coordinate $x$ coinciding with the direction of the gradient of the mixture fraction, the strain rate $a=-\frac{\partial u}{\partial x}$ is defined in terms of the gas velocity in the direction normal to the mixture fraction iso-surfaces, $u$~\cite{Olguin19}.

In Eqs.~\eqref{eq:Yk},~\eqref{eq:T} and~\eqref{eq:gz}, apart from the typical parameter $a$, the only unclosed quantities are the mass and energy evaporation source terms. In the literature, these have been closed by means of steady profiles obtained either by assuming a simplified evaporation model~\cite{Franzelli15} or by projecting physical space solutions into composition space~\cite{Olguin19}. As it has been pointed out before, these approaches exclude a big region of the possible solution space of the spray flamelet equations and this will be relaxed in the present work. 


\subsection{Closure of the Evaporation Source Terms}\label{sec:closure}

For the closure of the liquid phase, we adopt the Lagrangian approach originally introduced in~\cite{Continillo90} as an appropriate physical space formulation. This model assumes a dilute spray consisting of spherically symmetric, single component droplets and it individually tracks different droplet groups injected in selected time steps. The model neglects droplet–droplet interactions, coalescence and break-up but, while the different size groups are always monodisperse, it still allows for local polydispersity generated by their coexistence in a given numerical cell. With the information obtained from the individually tracked $J$ droplet size groups, the local mass and energy evaporation source terms are computed as
\begin{align}\label{eq:sv}
\dot{S}_v=\sum_{j=1}^Jn_j\dot{m}_j
\end{align}
and
\begin{align}\label{eq:se}
\dot{S}_e=\sum_{j=1}^J-n_j\left(\dot{q}_j+\dot{m}_j\left[\int_T^{T_{js}} C_{pF}dT+L_v\right]\right),
\end{align}
respectively. In Eqs.~\eqref{eq:sv} and~\eqref{eq:se}, $n$ represents the droplet number density, $\dot{m}$ is the evaporation rate, $\dot{q}$ is the energy flux into the droplet, $T_s$ is the temperature at the droplet surface and $L_v$ is the latent heat of vaporization. As readily noted, calculating these quantities requires consideration of the droplet dynamics, so that their exact location and surrounding local conditions are known. Since all elements within a droplet group share the same characteristics, calculations are performed for a single representative droplet and, for clarity, the subindex $j$ will be omitted in the remainder of this section.


We proceed now to introduce the different expressions required for the closure of $\dot{S}_v$ and $\dot{S}_e$. The mass vaporization rate, $\dot{m}$, can be calculated as~\cite{Continillo90,Abramzon89}
\begin{equation}\label{eq:dotm}
\dot{m}=2\pi \rho_fD_fR\widetilde{\mathrm{Sh}} \ln(1+\mathrm{B}_M),
\end{equation}
where the subscript $f$ refers to film properties, and $R$ is the instantaneous droplet radius. Here, the modified Sherwood number can be obtained from the following expression~\cite{Continillo90}
\begin{equation}
\widetilde{\mathrm{Sh}}=2+\frac{([1+\mathrm{Re}\mathrm{Sc}]^{1/3}[\max(1,\mathrm{Re})]^{0.077}-1)}{(1+\mathrm{B}_M)^{0.7}\log(1+\mathrm{B}_M)}\mathrm{B}_M,
\end{equation}
with $\mathrm{Re}$ and $\mathrm{Sc}$ denoting the Reynolds and Schmidt numbers, respectively. Further, the Spalding mass transfer number yields~\cite{Continillo90,Abramzon89}
\begin{equation}
\mathrm{B}_M=\frac{Y_{Fs}-Y_F}{1-Y_{Fs}},
\end{equation}
where the subscript $s$ indicates that the property is evaluated at the droplet surface. 

Alternatively, an expression for the mass vaporization rate, $\dot{m}$, can be obtained by relating it to the change of the droplet radius with time (or, equivalently, the droplet mass). This leads to~\cite{Continillo90,Abramzon89}
\begin{align}\label{eq:R}
\frac{d}{d\tau_l}\left(\frac{4}{3}\pi R^3\rho_l\right)=-\dot{m},
\end{align}
where the subscript $l$ refers to the liquid phase, and $\tau_l$ is the corresponding Lagrangian time (following the droplet). Inserting Eq.~\eqref{eq:dotm} into Eq.~\eqref{eq:R} allows computing the evolution of $R$ in time.

The energy flux into the droplet, $\dot{q}$, is calculated by the following expression~\cite{Continillo90,Abramzon89}
\begin{equation}\label{eq:qdot}
\dot{q}=\dot{m}\left(\frac{C_{pf}(T-T_{s})}{B_T}-L_v\right),
\end{equation}
where the Spalding energy transfer number is computed as~\cite{Continillo90,Abramzon89}
\begin{equation}
\mathrm{B}_T=(1-\mathrm{B}_M)^\phi-1,
\end{equation}
\begin{equation}
\phi=\frac{C_{pl}\widetilde{\mathrm{Sh}}}{C_{pf}\widetilde{\mathrm{Nu}}}\frac{1}{\mathrm{Le}}.
\end{equation}
The modified Nusselt number is computed from
\begin{equation}
\widetilde{\mathrm{Nu}}=2+\frac{([1+\mathrm{Re}\mathrm{Pr}]^{1/3}[\max(1,\mathrm{Re})]^{0.077}-1)}{(1+\mathrm{B}_T)^{0.7}\log(1+\mathrm{B}_T)}\mathrm{B}_T,
\end{equation}
where $\mathrm{Pr}$ denotes the Prandtl number. Finally, the temperature distribution inside the droplet is described through the conduction limit model as
\begin{align}\label{eq:ener_drop}~\cite{Continillo90,Abramzon89}
\frac{\partial T_l}{\partial \tau_l}=\frac{1}{r^2}\frac{\partial}{\partial r}\left(\alpha_lr^2\frac{\partial T_l}{\partial r}\right),
\end{align}
where $\alpha_l$ denotes the thermal conductivity of the liquid phase, $r$ is the internal radial coordinate, and $T_l$ is the droplet temperature, with $T_s=T_l(r=R)$. The boundary condition for  Eq.~\eqref{eq:ener_drop}, corresponds to the heat flux computed from Eq.~\eqref{eq:qdot} at $r=R$.


An analytical expression can be obtained for the droplet number density by means of a Lagrangian analysis similar to the one carried out in~\cite{Continillo90} for spray flames in physical space. In particular, assuming the droplets move exclusively along the flamelet, the total number of droplets per unit length within a droplet group, $N$, should remain constant throughout the evaporation process. Therefore, the following condition needs to be satisfied
\begin{align}\label{eq:N}
\frac{dN}{d\tau_l}=0.
\end{align}
Integrating and rewriting this equation in terms of $n$, we obtain
\begin{align}\label{eq:N_int}
\frac{n(\tau_l)\Delta Z}{g_Z(\tau_l)}=\frac{n_0\Delta Z_0}{g_{Z,0}},
\end{align}
where $\Delta Z$ is length of the cell in composition space and the subscript $0$ refers to the initial conditions. Rearranging Eq.~(\ref{eq:N_int}) leads to the following final analytical expression
\begin{align}
n(\tau_l)=\frac{n_0\Delta Z_0}{g_{Z,0}}\frac{g_Z(\tau_l)}{\Delta Z}.
\end{align}


Finally, since the gradient of the mixture fraction has been assumed to be aligned with the physical coordinate $x$, the following droplet motion equation in physical space can be adopted~\cite{Continillo90,Abramzon89}
\begin{align}\label{eq:drop_dyn}
\frac{4}{3}\pi R^3\rho_l\frac{du_l}{d\tau_l}=\frac{\pi R^2\rho}{2}(u-u_l)|u-u_l|C_D,
\end{align}
where the gravitational force has been neglected. Here, $u_l$ is the droplet velocity, and $C_D=\frac{12\nu}{|u-u_l|R}$ is the drag coefficient (calculated assuming Stokes flow) and $\nu$ is the kinematic viscosity of the gas (see~\cite{Continillo90}). 

Equation~\eqref{eq:drop_dyn} can be transformed to describe the dynamic in composition space ($Z$-space) using the following relation
\begin{equation}
u_l=\frac{dx_l}{d\tau_l}=\frac{1}{g_Z}\frac{dZ_l}{d\tau_l},
\end{equation}
which, after rearrangement, yields
\begin{align}
\frac{d^2Z_l}{d\tau_l^2} - \frac{1}{g_Z}\frac{dg_Z}{dZ}\left(\frac{dZ_l}{d\tau_l}\right)^2 + \frac{9\nu\rho}{2R^2\rho_l}\frac{dZ_l}{d\tau_l}=\frac{9\nu\rho g_Z}{2R^2\rho_l}u.\label{eq:zd}
\end{align}

Furthermore, by expressing the strain rate as $a=g_Z\frac{\partial u}{\partial Z}$ and integrating, the gas velocity can be computed directly as
\begin{align}\label{eq:drop_dyn_z}
u=u(Z=0)-\int_{Z=0}^Z\frac{a}{g_Z}dZ,
\end{align}
where $u(Z=0)$ is a boundary condition that needs to be specified.

Making use of Eq.~\eqref{eq:zd}, the droplet position can be determined at every time, so that the properties needed to compute $\dot{m}$, $\dot{q}$ and $C_{pF}$ can be readily calculated. With this, the evaporation related source terms appearing in the spray flamelet equations, Eqs.~\eqref{eq:sv} and~\eqref{eq:se} have been properly closed. 


\subsection{Solution Strategy}\label{sec:solstrat}

\begin{figure}[b!]
\centering
\includegraphics[width=0.48\textwidth]{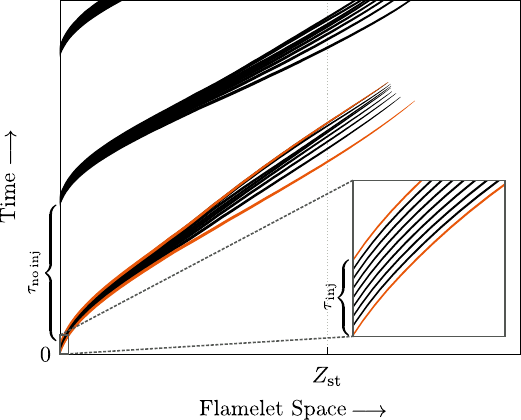}
\caption{Schematic of the pathways of droplet size groups dynamic in flamelet space over time. The thickness of the lines represents the instantaneous droplet radius.}
\label{fig1}
\end{figure}

The closed system of equations presented in the previous sections is solved using an in-house FORTRAN solver. Starting from the initial conditions for chemical species, temperature, and gradient of the mixture fraction, the liquid phase equations are solved first, followed by the computation of the source terms required to solve the gas phase. Due to the non-linearity of the problem, the gas phase equations are addressed iteratively, using a pseudo-transient time-stepping procedure until convergence is achieved. The simulation then advances to the next time step, continuing until either a steady or a quasi-steady state (cf. Section~\ref{sec:steady_unsteady}) is reached.

The liquid phase dynamics is calculated through multiple Lagrangian time steps, $d\tau_l$, during each gas-phase Eulerian time step, $d\tau$. For this, $d\tau_l$ is set smaller than $d\tau$ so that an accurate description is ensured. Additionally, injection can be deactivated at any time based on user-defined requirements for the cases with discontinuous injection.

In order to illustrate the kind of solutions obtained by our approach, Fig.~\ref{fig1} presents a schematic of the pathways of the droplets for a discontinuous injection case. As the lines move along the abscissa, they indicate that the droplet size groups penetrate further into the flamelet. A first group of droplets is injected at $\tau=0$, followed by subsequent injections over an interval $\tau_{\mathrm{inj}}$. Afterwards, the injection is paused for an interval $\tau_{\mathrm{no\,inj}}$. When new groups are injected, they coexist with the previous ones. Each group is tracked individually, which can lead to differences in the dynamics among groups injected within the same interval $\tau_{\mathrm{inj}}$ (orange lines) because the profiles of mass fractions and temperature are influenced by previously injected droplet groups.


\section{Results}\label{sec:results}

We present and analyze now different numerical solutions of the spray flamelet equations. For this, ethanol/air gas flamelets perturbed by different sprays are considered, which are established in a composition space ranging from $Z=0$ (pure air) to $Z=1$ (pure fuel), with a temperature of $300$~K at the air side and $360$~K at the fuel boundary. A detailed chemical reaction mechanism consisting of 38 species and 337 reactions is adopted~\cite{Chevalier93,Gutheil01}. The analysis includes the study of the flamelet structure for three specific cases (Section~\ref{sec:steady_unsteady}) and the generalization of the observations by means of a comprehensive parametric analysis of the effects of varying initial droplet radius and velocity (Section~\ref{sec:par_an}).


\subsection{Influence of the droplet injection strategy on flamelet structures}\label{sec:steady_unsteady}

It is shown now how different droplet injection strategies lead to very different solutions of the spray flamelet equations. For this, three cases will be analyzed in this section. The first of them, which is established as a reference and labeled as C$_0$, consists of a steady non-premixed flamelet perturbed by a spray injected from $Z=0$, which is generated through a continuous droplet injection. This case is equivalent to the flamelet structures analyzed in previous studies, where a steady evaporation profile computed in physical space was directly projected into mixture fraction space~\cite{Franzelli15, Olguin19}. The second and third cases, C$_1$ and C$_2$, correspond to two flames established by the adoption of a discontinuous droplet injection strategy, which leads to unsteady evaporation profiles. As explained before, the discontinuous droplet injection is characterized by the time periods of injection and non-injection of droplets, $\tau_{\mathrm{inj}}$ and $\tau_{\mathrm{no\,inj}}$, respectively. For both  C$_1$ and C$_2$, $\tau_{\mathrm{inj}}$ is selected as $0.05$~ms, while $\tau_{\mathrm{no\,inj}}$ is set to $0.1$ and $0.45$~ms, respectively. To ensure the same time-averaged liquid-fuel to air equivalence ratio for all these flames, the following condition is imposed at the injection point 
\begin{equation}
\bar{E}=\frac{\int E(\tau)d\tau}{\int d\tau}=0.1.
\end{equation}
While this is a rather moderate amount of liquid fuel, the effects of its addition are important, as will be shown in the next section. For the three cases the same initial droplet radius ($10$~\textmu m) and initial droplet velocity ($0.5$~m/s) are considered.

\begin{figure}[b!]
\centering
\includegraphics[width=1.\textwidth]{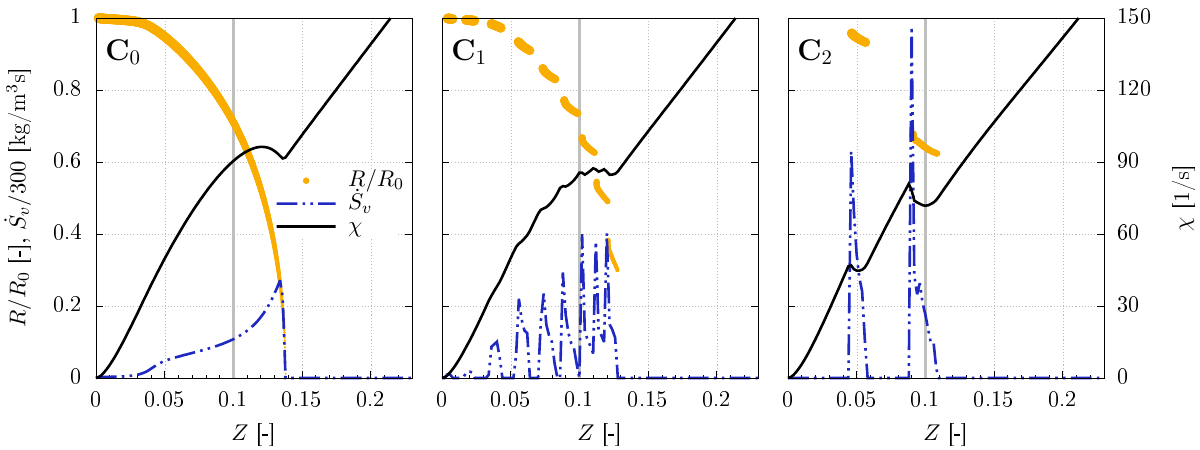}
\caption{Profiles of $R/R_0$, $\dot{S}_v$ and $\chi$ for flamelets C$_0$, C$_1$ and C$_2$. The vertical gray line represents $Z_\mathrm{st}\approx0.1$.}
\label{fig2}
\end{figure}

\begin{figure*}[t!]
\centering
\includegraphics[width=0.48\textwidth]{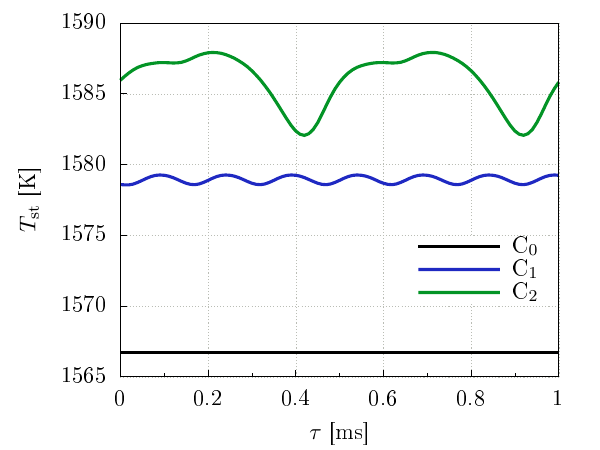}
\includegraphics[width=0.48\textwidth]{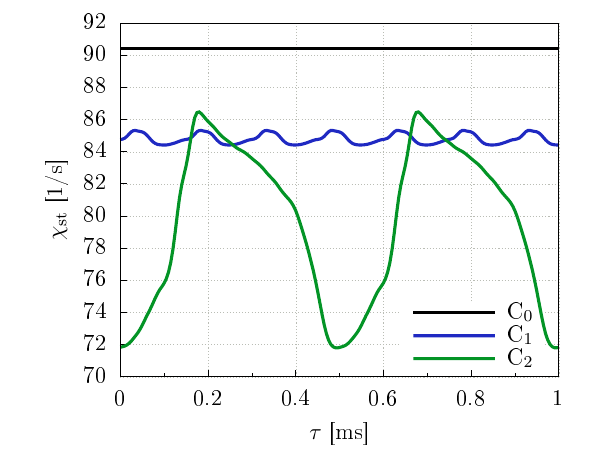}
\caption{Temperature (left) and scalar dissipation rate (right) at stoichiometry, over time for flamelets C$_0$, C$_1$ and C$_2$.}
\label{fig3}
\end{figure*}

\begin{figure*}[t!]
\centering
\includegraphics[width=1.\textwidth]{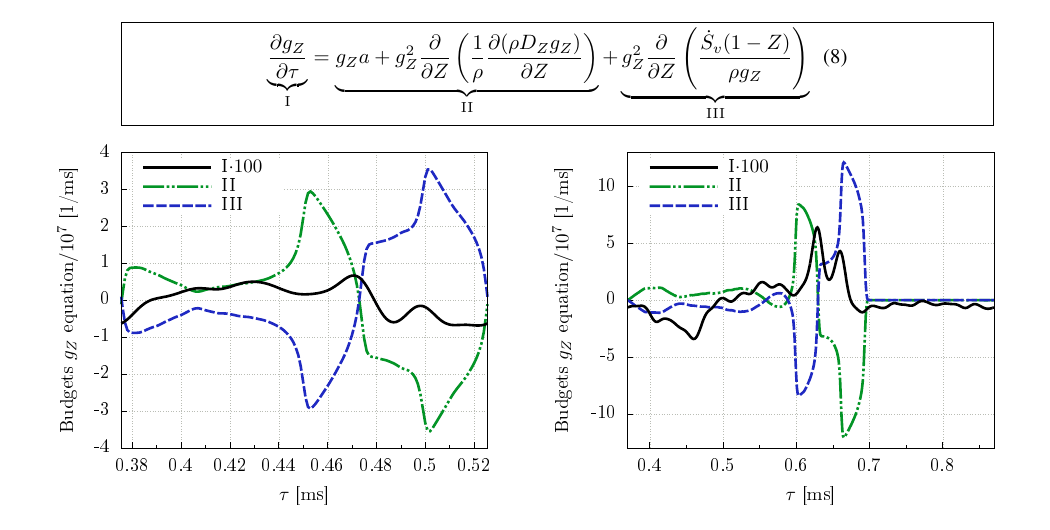}
\caption{Budgets of the $g_Z$ equation, Eq.~\eqref{eq:gz}, for flamelets C$_1$ (left), and C$_2$ (right) over time at stoichiometric point. For the readers' convenience the equation is shown at the top of the figure.}
\label{fig4}
\end{figure*}

\begin{figure*}[t!]
\centering
\includegraphics[width=0.48\textwidth]{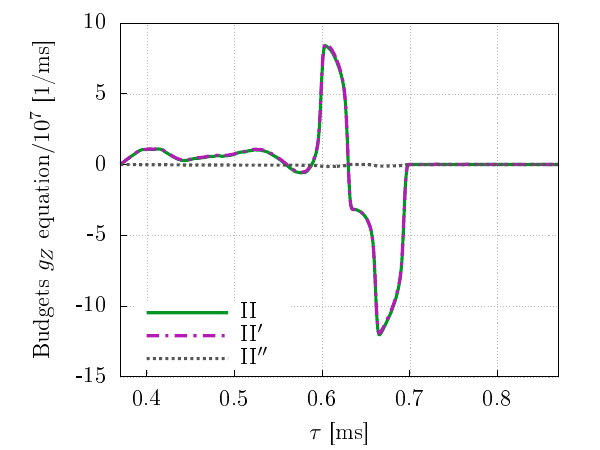}
\includegraphics[width=0.48\textwidth]{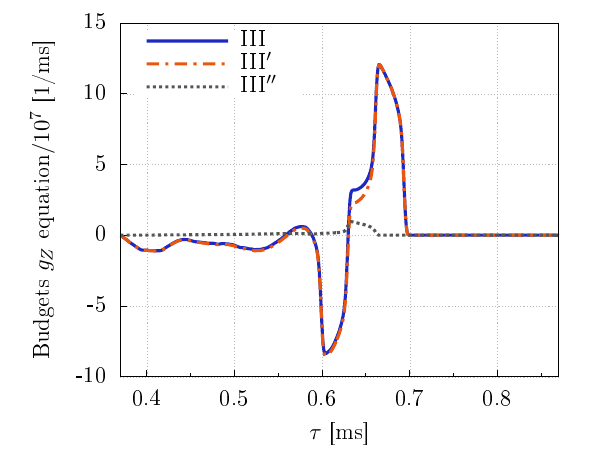}
\caption{Budgets of the decomposition of the strain and diffusion terms, Eq.~\eqref{eq:diff-deco} (left), and the evaporation-related term, Eq.~\eqref{eq:evap-deco} (right), for flamelet C$_2$ over time at stoichiometric point.}
\label{fig5}
\end{figure*}

In Fig.~\ref{fig2}, profiles of the normalized instantaneous droplet radius, $R/R_0$, the evaporation source term, $\dot{S}_v$, and the scalar dissipation rate, $\chi$, are presented for C$_0$, C$_1$ and C$_2$, respectively. It is clear from these figures that, even when the same set of spray parameters is considered (initial droplet radius, velocity and equivalence ratio), the specific droplet injection strategy adopted leads to significantly different evaporation and scalar dissipation rate profiles. First, as it has been previously pointed out, C$_0$ corresponds to a steady flame, while for cases C$_1$ and C$_2$ the evaporation is inherently unsteady, which introduces important oscillations in other quantities (profiles shown in Fig.~\ref{fig2} are instantaneous snapshots). Further, it is observed that the different local peaks appearing in the evaporation profiles generate corresponding local maximum values of the scalar dissipation rate. This, as it will be shown next, strongly modifies the local temperature and, consequently, the flamelet stability.

The evolution of the temperature and the scalar dissipation rate at $Z_\mathrm{st}$, $T_\mathrm{st}$ and $\chi_\mathrm{st}$, respectively, are displayed in Fig.~\ref{fig3} for the three cases under consideration. Here, the quasi-steady nature of cases C$_1$ and C$_2$ is clearly appreciated, with the respective oscillations matching the injection and non-injection cycles of the droplets ($0.15$ and $0.5$~ms, respectively). Significant differences are observed for $T_\mathrm{st}$, with the quasi-steady flamelets having higher temperatures than C$_0$. This increase in $T_\mathrm{st}$ can be directly related to the decrease of the mean value of $\chi_\mathrm{st}$, which indicates shorter residence times of the chemical species within the reaction zone~\cite{Peters84}.
 
In order to illustrate the importance of evaporation (and therefore the droplet injection strategy) in the definition of the scalar dissipation rate profile, the budgets of the $g_Z$-equation evaluated at $Z_\mathrm{st}$ are shown in Fig.~\ref{fig4} for a completely cycle of their quasi-steady evolution. Here, term I corresponds to the transient effects (amplified by a factor of $100$). Term II, on the other hand, comprises the strain and diffusion effects, which correspond to the gas phase contributions. Finally, term III represents the effects of the perturbation introduced by the spray. While the duration of the analyzed cycles differs, the behavior of the different terms of the $g_Z$-equation are qualitatively the same for both cases. More specifically, it is readily observed that the cycles can be split into to parts. During the first part of the cycle, term III is negative, while term II tries to compensate it. During the second part of it, the signs of these two terms rapidly switch due to a phenomenon that cannot be observed in a steady flame such as the one selected for case C$_0$. As it will be shown next, this transition can be directly related to the changes of sign of the gradient of evaporation and the second derivative of $g_Z$ (gradient dissipation). For this, terms II and III are rewritten as
\begin{align}\label{eq:diff-deco}
&\underbrace{ag_Z+g_Z^2\frac{\partial}{\partial Z}\left(\frac{1}{\rho}\frac{\partial(\rho D_Zg_Z)}{\partial Z}\right)}_{\mathrm{II}} =  \underbrace{g_Z^2D_Z\frac{\partial^2g_Z}{\partial Z^2}}_{\mathrm{II}^\prime}
\underbrace{+ag_Z+g_Z^2\left(\frac{\partial}{\partial Z}\left(\frac{g_Z}{\rho}\frac{\partial (\rho D_Z)}{\partial Z}\right)+\frac{\partial D_Z}{\partial Z}\frac{\partial g_Z}{\partial Z}\right)}_{\mathrm{II}^{\prime\prime}}
\end{align}
and
\begin{align}\label{eq:evap-deco}
& \underbrace{g_Z^2\frac{\partial}{\partial Z}\left(\frac{\dot{S}_v(1-Z)}{\rho g_Z}\right)}_{\mathrm{III}} = 
+\underbrace{\frac{g_Z}{\rho}\frac{\partial(\dot{S}_v(1-Z))}{\partial Z}}_{\mathrm{III}^\prime}
\underbrace{- \left(\frac{1}{g_Z}\frac{\partial g_Z}{\partial Z}+\frac{1}{\rho}\frac{\partial\rho}{\partial Z}\right)\frac{\dot{S}_v(1-Z)}{\rho g_Z}}_{\mathrm{III}^{\prime\prime}},
\end{align}
respectively. 

\begin{figure}[t!]
\centering
\includegraphics[width=0.48\textwidth]{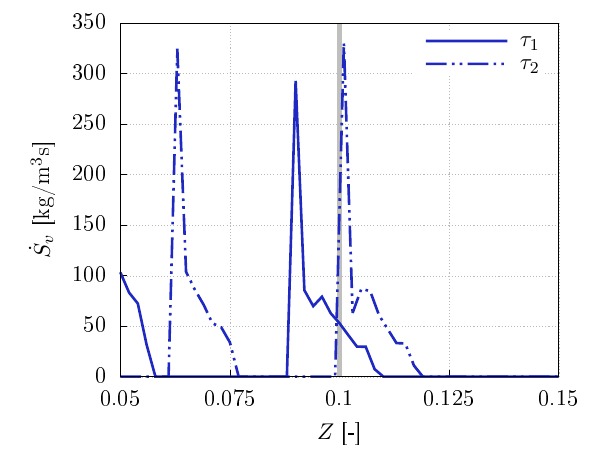}
\caption{Profiles of $\dot{S}_v$ at two different time instants: $\tau_1=0.5$~ms, before the peak reaches the stoichiometric point (continuous line), and $\tau_2=0.65$~ms, after it passed stoichiometry (dashed line).}
\label{fig6}
\end{figure} 

The budgets of Eqs.~\eqref{eq:diff-deco} and~\eqref{eq:evap-deco} are shown in Fig.~\ref{fig5} for case C$_2$, where it can be seen that II$^{\prime}$ and III$^{\prime}$ are the only components with non-negligible contributions to the $g_Z$-equation budgets. Additionally, since $g_Z$, and $\rho$ cannot be negative, the signs of II and III are completely determined by $\frac{\partial^2 g_Z}{\partial Z^2}$ and $\frac{\partial \dot{S}_v}{\partial Z}$, respectively. The change of signs of these quantities are determined by the liquid phase dynamics as follows: When a group of droplets approaches the stoichiometric point, the local gradient of $\dot{S}_v$ is negative and when the droplet group already passed, it is positive (see Fig.~\ref{fig6}). Since the peaks of $g_Z$ coincide with the peaks of evaporation, and given that $g_Z$ dissipation occurs from the peaks towards the left and right sides, the same change of sign is observed for this quantity.

The results presented in this section clearly show the importance of the droplet injection strategy, which can lead to highly unsteady evaporation profiles significantly enhancing temperatures, despite the quantity of liquid fuel considered in these numerical experiments is relatively low ($E=0.1$ for all analyzed cases). Additionally, the budget analysis of the $g_Z$-equation shows that the terms associated with the gas phase (term II) and the liquid phase (term III) are of comparable magnitude. In the next section, it will be shown that the mentioned effect of discontinuous injection is of major importance on flame stability. For this, the analysis carried out here for specific conditions only, will be extended through a parametric analysis of the effects of varying $u_{0}$ and $R_{0}$.


\subsection{On the generality of the findings}\label{sec:par_an}

\begin{figure}[b!]
\centering
\includegraphics[width=0.48\textwidth]{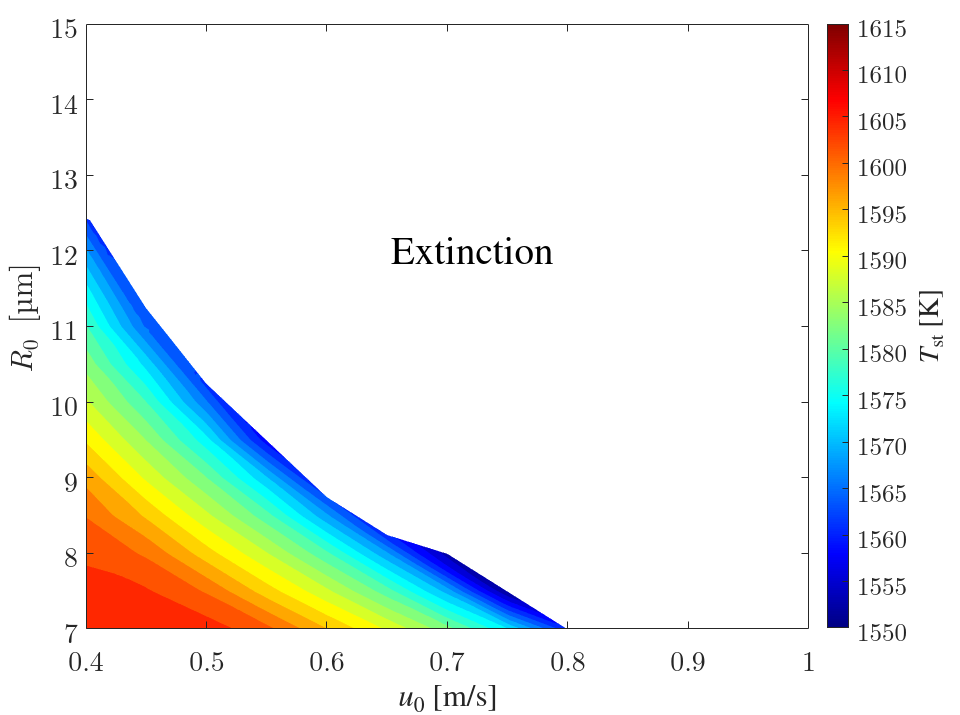}
\includegraphics[width=0.48\textwidth]{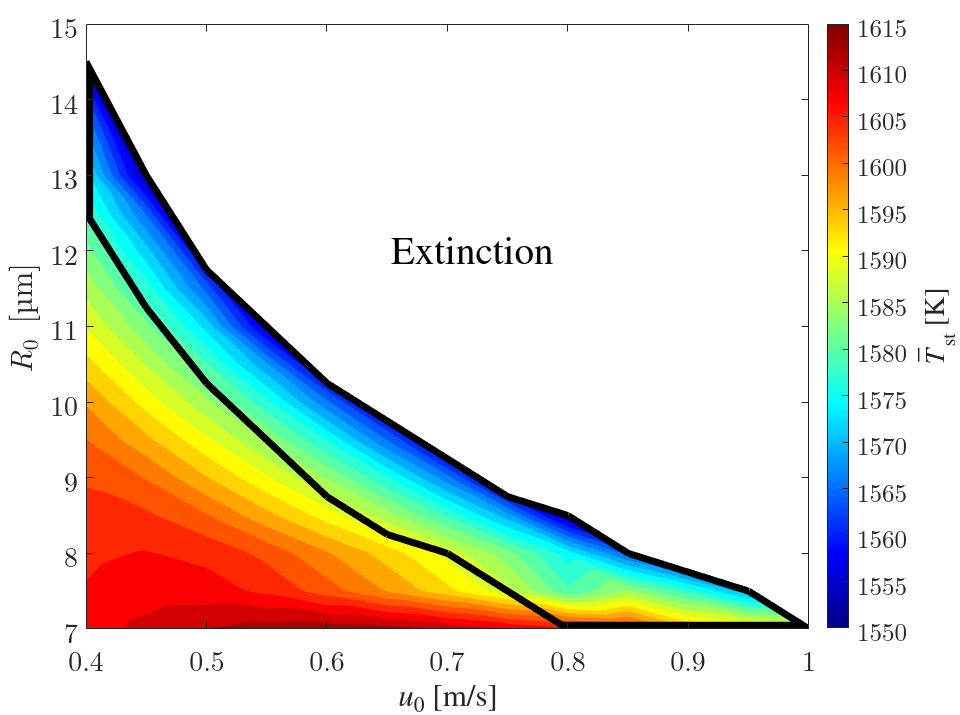}
\caption{$T_\mathrm{st}$ for extended C$_0$ (top) and extended C$_2$ (bottom) flamelets, varying its initial droplet velocity and radius. The white areas indicate extinguished flamelets.}
\label{fig7}
\end{figure}

In the previous section, it was shown how a discontinuous droplet injection strategy led to higher flamelet temperatures than an equivalent continuous one. The numerical experiments, however, were conducted for a single set of spray parameters and, in this section, the results are generalized through a parametric analysis of the effects of varying the initial droplet radius and velocity. For this, the conditions specified for cases C$_0$ and C$_2$ will be systematically modified to cover values of $u_0$ between $0.4$ and $1$~m/s and of $R_0$ between $7$ and  $15$~\textmu m. In the remainder of this work, we will refer to these modified experiments as extended C$_0$ and extended C$_2$, respectively. To facilitate the comparison between steady and quasi-steady flamelet structures, the mean temperature at stoichiometry of the latter will be calculated as 
\begin{equation}
\bar{T}_\mathrm{st}=\frac{\int T_\mathrm{st}(\tau)\, d\tau}{\int d\tau}.
\end{equation} 
    
Figure~\ref{fig7} displays a comparison of the results obtained with both injection strategies, with the white areas representing extinguished flamelets. It is observed that for both extended C$_0$ and C$_2$, there is a high temperature region located at low values of the initial droplet radius and velocity. This can be explained by the low droplet penetration and the corresponding early droplet evaporation of the spray. In general, a high droplet penetration is associated with a low droplet residence time within the reaction zone which does not favor combustion~\cite{Ying22}.

Despite the similarities, two major differences can be observed between the extended C$_0$ and C$_2$ experiments. First, the range of $u_0$-$R_0$ combinations leading to burning flamelets is considerably larger for the latter (see area within the black line in the bottom of Fig.~\ref{fig7}). Secondly, for the conditions where both steady and unsteady evaporation profiles can keep the flame burning, the adoption of a discontinuous droplet strategy consistently leads to higher $T_\mathrm{st}$ values.

In summary, the analysis presented in this section confirms the generality of the findings reported in Section~\ref{sec:steady_unsteady}. The extended region of $u_0$-$R_0$ combinations leading to burning flamelets represents additional support of the conclusion that, provided the spray parameters are properly selected, discontinuous injection strategies can considerably improve flamelet stability. 


\subsection{Further topics of interest: Extinction and re-ignition phenomena}\label{sec:ext_reig}

As already pointed out, this work aims to provide an appropriate mathematical basis for the analysis of important combustion situations that have not been studied so far. Among the phenomena fitting this definition are extinction and re-ignition processes, which will be briefly addressed now for illustrative purposes. More specifically, a numerical experiment is carried out, which is very similar to the one reported by Mau\ss~et al.~\cite{Mauss91}. The analysis starts with the flamelet C$_2$ at $a=850$/s, which is then increased to $985$/s for the duration of a single injection-non-injection cycle ($0.5$~ms). After that, $a$ is decreased back to $850$/s. Two variations of this procedure are considered: i) In flamelet C$_\mathrm{2,A}$, the increase in $a$ is imposed when $\chi_\mathrm{st}$ reaches its maximum value, whereas ii) in flamelet C$_\mathrm{2,B}$, the modification is introduced when $\chi_\mathrm{st}$ is at its minimum.

\begin{figure}[b!]
\centering
\includegraphics[width=.48\textwidth]{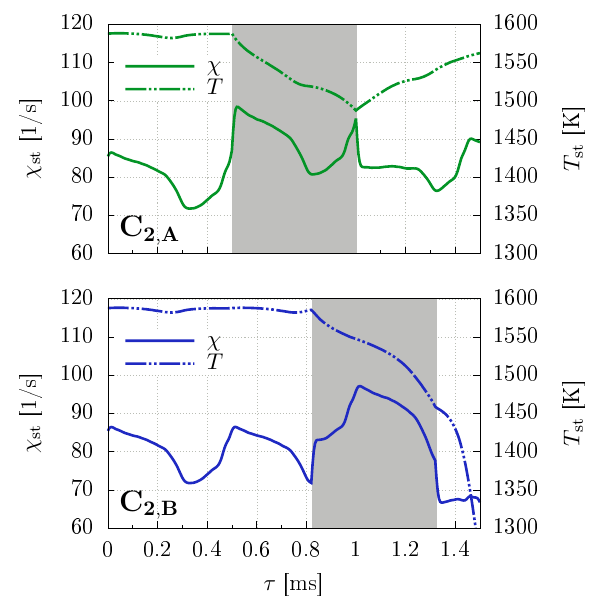}
\caption{Scalar dissipation rate (continuous line) and temperature (dashed line) at stoichiometry, over time for flamelets C$_\mathrm{2,A}$ (top) and C$_\mathrm{2,B}$ (bottom), under a strain rate increase from $850$/s to $985$/s for $0.5$~ms. The gray area denotes the interval with high strain rate.}
\label{fig8}
\end{figure}

In Fig.~\ref{fig8}, profiles of $\chi_\mathrm{st}$ and $T_\mathrm{st}$ are displayed as a function of time, with the period in which $a$ is increased indicated by the gray area. Interestingly, even though both C$_\mathrm{2,A}$ and C$_\mathrm{2,B}$ correspond to perturbations of the same flamelet, only the former is able to re-ignite after the strain rate reduction. In C$_\mathrm{2,A}$, since the increase in $a$ is applied when $\chi_\mathrm{st}$ reaches its maximum, it is overcome by the decreasing phase in the cyclic behavior of the instantaneous scalar dissipation rate (induced by evaporation, as discussed in Section~\ref{sec:steady_unsteady}). This results in a reduction of $\chi$, thus mitigating the decrease in $T$. Conversely, in C$_\mathrm{2,B}$, the increasing phase of the scalar dissipation rate in the cycle and the strain rate effect coincide. Thus, they keep the high value of $\chi$ longer, leading to a much stronger temperature reduction that prevents the flame from re-igniting after the subsequent decrease of $a$.

This simple experiment illustrates the sort of phenomena that have not yet been studied in the literature, although similar strategies have been employed in unsteady gas flamelets~\cite{Paxton19}, and for which the current approach provides a very suitable framework. 

\section{Conclusions}\label{sec:con}

So far, solutions of the spray flamelet equations analyzed in the literature have focused on situations presenting steady evaporation profiles only. In this work, these equations were solved to analyze the effects of unsteady evaporation profiles on non-premixed ethanol/air flamelets. For this, a Lagrangian description of the liquid phase in composition space was adopted, which allows the consideration of discontinuous droplet injection. The results show that the intrinsically unsteady interactions introduced by this injection strategy can lead to enhanced temperatures and reduced scalar dissipation rates at stoichiometry. These findings have been generalized by means of parametric analyses of the influence of initial droplet radius and velocity, which confirmed their validity for a wide combination of these variables. These numerical experiments have also shown that discontinuous droplet injection considerably increases the combination of initial droplet parameters ($u_0$ and $R_0$) under which burning flamelets can be obtained, a consequence of the improved stability (higher temperatures and lower scalar dissipation rates) achieved by means of this injection strategy. 

The proposed formulation opens a wide spectrum of possibilities in terms of the solution of these equations that can be effectively studied. This has been briefly illustrated by means of an analysis of extinction and re-ignition of a flamelet. 


\section*{Acknowledgments}

The authors acknowledge funding from Proyecto Interno USM PI\_LIR\_2022\_15, FH and FR thank ANID (Chile) for financial support through the Doctorate Scholarships 21201826 and 21212269, respectively. FH is also grateful for the financial contribution from the PIIC-USM grant. AS and CH acknowledge funding by the Deutsche Forschungsgemeinschaft (DFG, German Research Foundation) - Projektnummer 325144795.

\bibliography{references}

\end{document}